\documentclass{article}
\usepackage{spconf,amsmath,graphicx}

\usepackage{amsfonts,color}
\usepackage{xcolor}

\usepackage{hyperref}
\hypersetup{
    colorlinks=true,
    linkcolor=blue,
    filecolor=magenta,
    urlcolor=blue,
    citecolor=blue,
}

\let\OLDthebibliography\thebibliography
\renewcommand\thebibliography[1]{
  \OLDthebibliography{#1}
  \setlength{\parskip}{0pt}
  \setlength{\itemsep}{0pt plus 0.35ex}
}

\title{Hierarchical prosody modeling and control in non-autoregressive parallel neural TTS}

\name{Tuomo~Raitio\qquad Jiangchuan~Li\qquad Shreyas~Seshadri}

\address{Apple}

\begin{document}
\ninept

\maketitle

\begin{abstract}
Neural text-to-speech (TTS) synthesis can generate speech that is indistinguishable from natural speech. However, the synthetic speech often represents the average prosodic style of the database instead of having more versatile prosodic variation. Moreover, many models lack the ability to control the output prosody, which does not allow for different styles for the same text input. In this work, we train a non-autoregressive parallel neural TTS front-end model hierarchically conditioned on both coarse and fine-grained acoustic speech features to learn a latent prosody space with intuitive and meaningful dimensions. Experiments show that a non-autoregressive TTS model hierarchically conditioned on utterance-wise pitch, pitch range, duration, energy, and spectral tilt can effectively control each prosodic dimension, generate a wide variety of speaking styles, and provide word-wise emphasis control, while maintaining equal or better quality to the baseline model.
\end{abstract}


\section{Introduction}

A typical neural text-to-speech (TTS) system consists of two components: a neural front-end \cite{shen2017natural, ping2018deep, ren2020fastspeech2} that maps from character or phoneme input to intermediate features, such as Mel-spectrograms, and a neural back-end \cite{oord2016wavenet, kalchbrenner2018efficient} that maps from the Mel-spectrogram into a sequence of speech samples. These two networks are trained with a large amount of speech data, generating speech that can be indistinguishable from natural speech \cite{shen2017natural}.



Tacotron, an attention-based encoder-decoder model \cite{shen2017natural} is one of the most widely used neural TTS front-end architectures. However, there are several downsides with this type of model. First, the autoregressive architecture is inefficient both at training and inference time with modern parallel computing hardware. Second, due to the the attention-based alignment, the model suffers from instabilities at inference time, resulting in skipping or repetition of words and mumbling. Third, Tacotron uses teacher-forcing that introduces a discrepancy between training and inference.

Non-autoregressive parallel TTS architectures \cite{ren2019fastspeech,ren2020fastspeech2, elias2020parallel, donahue2021endtoend} are computationally more efficient and more stable, however, they need to solve two significant problems. First, they need a robust way of aligning the input and output sequences. Many recent non-autoregressive methods use an explicit duration model to get the alignment \cite{ren2019fastspeech, ren2020fastspeech2}. Secondly, they need to solve the one-to-many mapping problem, i.e., solve the information gap on how to realize the prosodic output of speech. While autoregressive models can use previously generated frames as context, non-autoregressive parallel methods need a different way to solve this. Conditioning on explicit prosodic features such as pitch and energy \cite{ren2020fastspeech2, chien2021hierarchical} or using latent variables \cite{Wang2018StyleTU, hsu2018hierarchical, zhang2019latent, sun2020fully, chien2021hierarchical} can be used to solve the problem.

Neural TTS front-end models tend to learn prosody as an inherent part of the model, however, such prosodically unsupervised models lack the explicit control over the output prosody. In recent parallel architectures, the prosody is more explicitly modeled to solve the one-to-many mapping \cite{ren2020fastspeech2, chien2021hierarchical}, however, they may still lack a useful control over the output prosody. While several methods have been proposed to model and control the output prosody in autoregressive models, such as using latent variables \cite{skerryryan2018endtoend, hsu2018hierarchical, zhang2019latent, sun2020fully}, style tokens \cite{Wang2018StyleTU}, or explicit prosodic features \cite{Shechtman_2019, raitio2020prosody}, few studies have investigated explicit prosody modeling and control in parallel architectures.

In this paper, we describe a non-autoregressive parallel TTS front-end architecture with hierarchical prosody modeling and control. The architecture is similar to FastSpeech 2 \cite{ren2020fastspeech2} with explicit phone-wise duration modeling and conditioning of the decoder with phone-wise pitch and energy. However, this kind of conditioning does not allow for higher-level modeling and control over the prosodic output as there is no simple method to modify the fine-grained features to achieve a desired prosody. Therefore, we train and predict utterance-wise (coarse-grained) prosodic features first, which are then used to condition the phone-wise (fine-grained) prosodic predictors. This enables hierarchical prosody modeling, and most importantly, higher-level control over the prosodic output.

Our main contribution in this work is a fast non-autoregressive parallel neural TTS front-end architecture with hierarchical prosody modeling and control using intuitive prosodic features. While there are several approaches that use hierarchical prosody modeling in autoregressive TTS \cite{ronanki2017, ribeiro2016, yin2016, hsu2018hierarchical, wan2019chive, sun2020fully, Shechtman2021, shechtman21_interspeech}, this study investigates hierarchical prosody modeling in parallel neural TTS. Hierarchical prosody modeling approaches in parallel TTS are presented in \cite{chien2021hierarchical, bae21_interspeech}, however, they aim to improve the overall quality rather than allow expressive control using higher-level features, as in \cite{Shechtman_2019, raitio2020prosody, Shechtman2021} and this study. Previously, we presented prosody modeling and control using Tacotron 2 \cite{raitio2020prosody}, and now we expand this work to hierarchical prosody modeling in non-autoregressive parallel TTS. We also extend our model to utilize the fourth dimension of prosody: spectral tilt \cite{Campbell2003vq, raitio2020prosody}. We show that each feature has a specific and effective control over the prosodic space, and that the proposed method can generate various speaking styles, and even provide word-wise emphasis control, while maintaining high quality\footnote{Speech samples can be found at \href{https://apple.github.io/parallel-tts-hierarchical-prosody-control/}{\nolinkurl{https://apple.github.io/parallel-tts-hierarchical-prosody-control/}}.}.

\section{Technical overview}

\begin{figure}[t]
  \centering
  \includegraphics[width=1.0\linewidth]{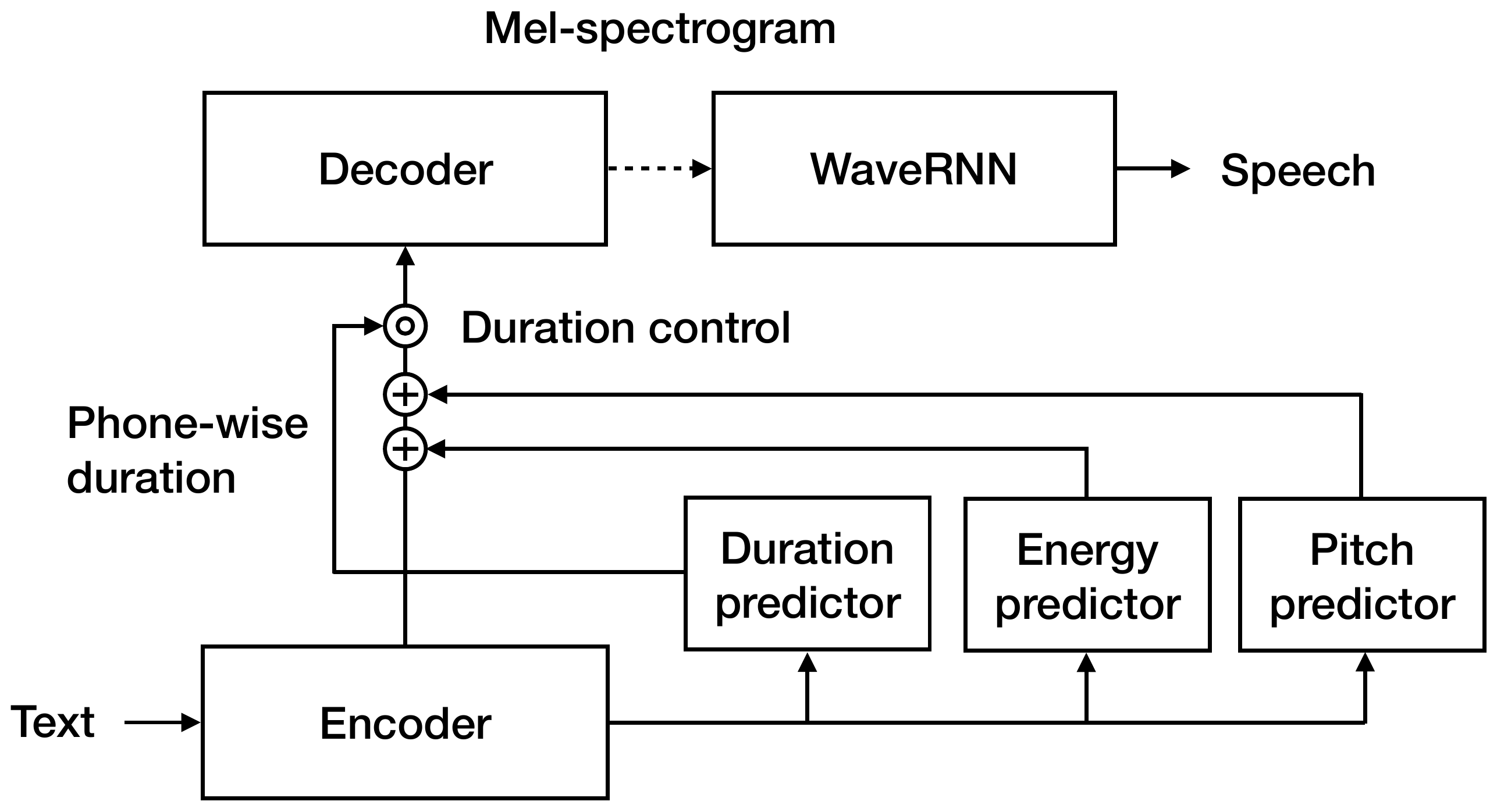}
  \vspace{-7mm}
  \caption{Baseline parallel TTS model front-end architecture.}
  \vspace{-5mm}
  \label{fig:architecture_baseline}
\end{figure}

The architecture of our baseline parallel TTS front-end model, illustrated in Fig.~\ref{fig:architecture_baseline}, is similar to FastSpeech 2 \cite{ren2020fastspeech2}. The input is a phoneme sequence with punctuation and word boundaries, and the output is a Mel-spectrogram. The model is based on a feed-forward Transformer (FFT) \cite{vaswani2017attention, ren2019fastspeech} encoder and dilated convolution decoder. The encoder consists of an embedding layer that converts the phoneme sequence to phoneme embeddings followed by a series of FFT blocks that take in the phoneme embeddings with positional encodings and output the phoneme encodings. Each FFT block consists of a self-attention layer \cite{vaswani2017attention} and 1-D convolution layers along with layer normalization and dropout. The phoneme encodings are then fed to the variance adaptors that predict phone-wise duration, pitch, and energy. The variance adaptors consist of 1-D convolution layers along with layer normalization and dropout similar to \cite{ren2020fastspeech2}. Instead of using pitch spectrograms as in \cite{ren2020fastspeech2}, we use continuous pitch, quantization, and finally projection to an embedding. The predicted phone-wise pitch and energy features are then added to the phoneme encodings, after which they are upsampled according to the predicted phone-wise durations. The decoder consists of a series of dilated convolution stacks instead of the original FFT blocks in \cite{ren2020fastspeech2}, which improves model inference speed as well as saving runtime memory compared with the original design. Finally, the decoder converts the adapted encoder sequence into a Mel-spectrogram sequence in parallel. The baseline model achieves 150x faster than real-time synthesis on a GPU and 100x on a mobile device.

To generate a sequence of speech samples from the Mel-spectrogram, we use an autoregressive recurrent neural network (RNN) based back-end architecture, similar to WaveRNN \cite{kalchbrenner2018efficient}. Although WaveRNN is an autoregressive model, it can be replaced with a parallel back-end model to achieve a fully parallel architecture. The model consists of a single RNN layer with 512 hidden units, conditioned on Mel-spectrogram, followed by two fully-connected layers ($512\times256$, $256\times256$), with single soft-max sampling at the output. The model is trained with pre-emphasized speech sampled at 24~kHz and $\mu$-law quantized to 8 bits for efficiency. We can run three models in parallel on a single GPU, each of them generating speech 7.7x faster than real-time, while our on-device implementation runs 3.3x faster than real-time on a mobile device. More information about the neural back-end implementation can be found in \cite{appleneuraltts2021asru}.

\begin{figure}[t]
  \centering
  \includegraphics[width=1.0\linewidth]{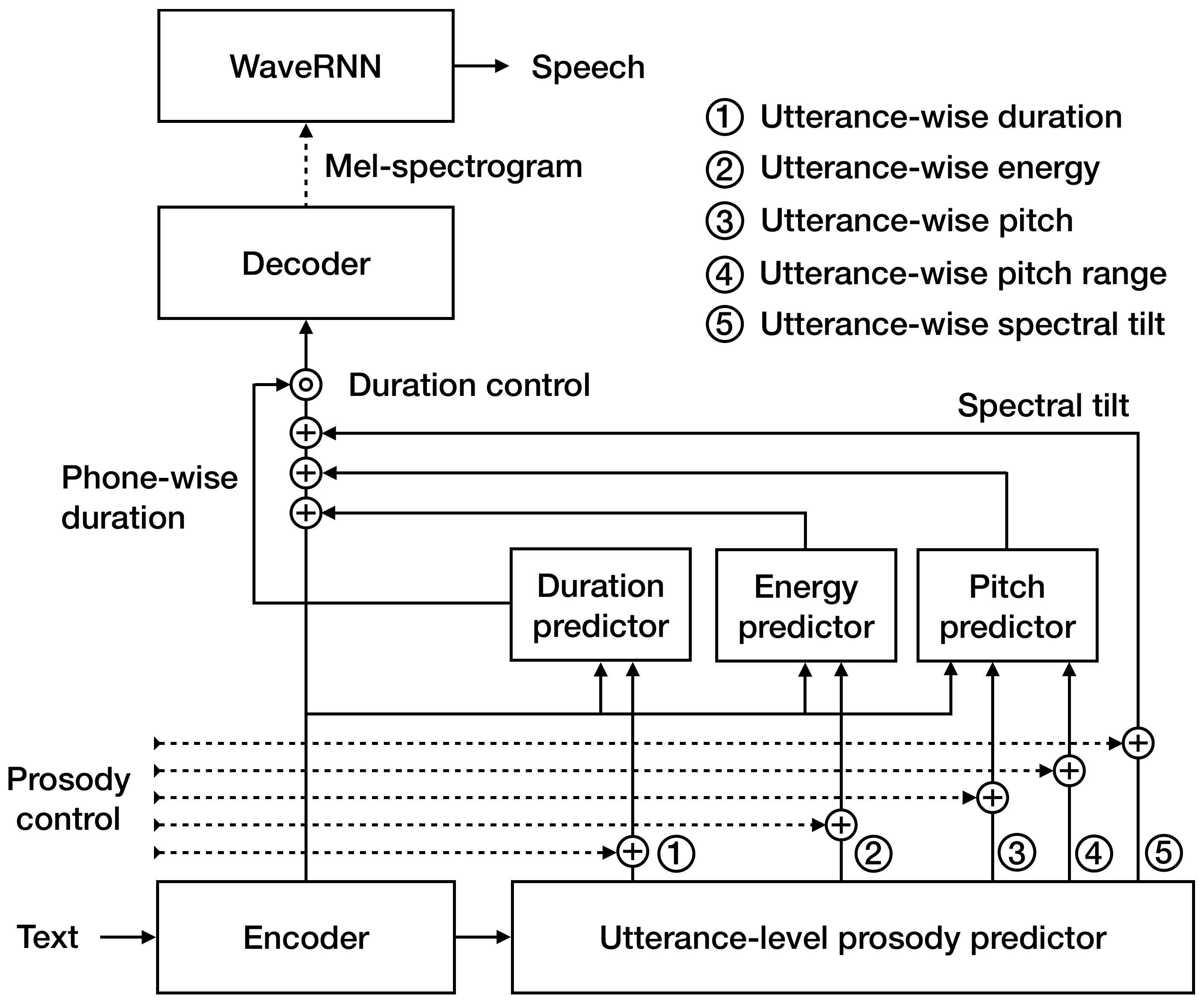}
  \vspace{-7mm}
  \caption{Proposed parallel TTS front-end model architecture with hierarchical prosody modeling and control.}
  \vspace{-5mm}
  \label{fig:architecture_proposed}
\end{figure}

\subsection{Proposed hierarchical prosody modeling architecture}

The proposed hierarchical prosody modeling front-end architecture is illustrated in Fig.~\ref{fig:architecture_proposed}. We extend the baseline architecture by introducing a new variance adaptor at utterance level. We predict the utterance-wise prosodic features (repeated at every phone in a sentence) from the encoder outputs, namely, pitch, pitch range, duration, energy, and spectral tilt (see Sec.~\ref{sec:prosodic_features} for details). The architecture of the new utterance-level variance adaptor is identical to the phone-level adaptors except that we predict all the five features using the same module (output dimension is 5 instead of 1). The output features are concatenated with the corresponding variance adaptors as follows: utterance-wise duration is concatenated to the input of the duration predictor; utterance-wise pitch and pitch range are concatenated to the input of the pitch predictor; utterance-wise energy is concatenated to the input of the energy predictor; and utterance-wise spectral tilt is concatenated directly with the encoder output since there is no separate predictor for fine-grained spectral tilt. We use teacher-forcing of both utterance-wise and phone-wise prosodic features to efficiently train the network. At synthesis time, higher-level prosody control is achieved by adding bias to the corresponding utterance-wise prosody predictions. Fine-grained prosody control can be achieved by directly modifying the phone-wise features (not depicted in the figure nor used in this study).

\subsection{Prosodic features}
\label{sec:prosodic_features}

To learn a meaningful latent space for prosody control, we use acoustic features extracted from the original speech to condition the model. We use pitch, phone duration, speech energy, and spectral tilt to model the prosodic space. These features are easy to calculate from speech signals and are robust against background noise or other recordings conditions. These features are also disentangled to a large degree so that they can be varied independently \cite{raitio2020prosody}. Overall, these features provide intuitive control over the prosodic space.

We extract the pitch of speech using 3 pitch estimators, and then vote for the most likely pitch trajectory. We extract the frame-wise log-energy of each utterance as $E = 20log10(\hat{x})$, where $\hat{x}$ is the average absolute sample amplitude, excluding silences. We calculate the frame-wise spectral tilt of voiced speech using the predictor coefficient of a first order all-pole filter. We use automatic speech recognition to force-align the text and audio to obtain the phone durations. The frame-wise pitch and energy features are aggregated per phone to obtain the fine-grained features used in both the baseline and proposed model. We compute two utterance-wise features from the voted pitch contour: average log-pitch and log-pitch range of voiced speech. The latter is calculated as the difference between the 0.05 and 0.95 quantile values of the frame-wise log-pitch contour for each utterance. We also calculate the average log-phone duration, average log-energy, and average spectral tilt per utterance. The 5 utterance-wise features, log-pitch, log-pitch range, log-phone duration, log-energy, and spectral tilt, are then normalized to $[-1,1]$ by first calculating the median ($M$) and the standard deviation ($\sigma$) of each feature, and then projecting the data in the range [$M-3\sigma$, $M+3\sigma$] into $[-1,1]$. Finally, we clip values $|x|>1$ so that all data is in the range $[-1,1]$. This process is the same as in \cite{raitio2020prosody}.

\section{Experiments}

\subsection{Data}
\label{sec:data}

We use data from two American English speakers, a high-pitched speaker with 36-hour dataset (Voice 1), and a low-pitched speaker with 23-hour dataset (Voice 2).

\subsection{Models}
\label{sec:models}

We trained TTS voices with the following models:

\vspace{-1mm}
\begin{itemize}
\setlength\itemsep{-0.1mm}

\item[1.] {\bf Compact:} Lower-quality statistical TTS model.
\item[2.] {\bf UnitSelection:} High-quality unit selection model \cite{siriusel}.
\item[3.] {\bf Tacotron2:} High-quality Tacotron 2-based model \cite{appleneuraltts2021asru}.
\item[4.] {\bf Parallel:} High-quality parallel TTS model.
\item[5.] {\bf HParallel:} Proposed hierarchical parallel TTS model.

\end{itemize}
\vspace{-1mm}
The {\it Compact} voice, trained with audio from a different high-pitched speaker with 22.05~kHz sampling rate, acts as a lower-quality anchor for the listening tests. The rest of the models (2--5) were trained with both speakers described in Sec.~\ref{sec:data}. The {\it UnitSelection} \cite{siriusel} voices use the original 48~kHz recordings while the rest of the neural models use audio down-sampled to 24~kHz for training. However, the models 3--5 include a post-processing module that performs a linear interpolation of the 24~kHz synthetic speech to 48~kHz to compensate for the lack of high frequencies, acting as a simple but effective bandwidth extension algorithm \cite{appleneuraltts2021asru}. For {\it Parallel} and {\it HParallel} models, we use phone-wise duration, pitch, and energy as the fine-grained features. For {\it HParallel} models, we use utterance-wise pitch, pitch range, phone duration, speech energy, and spectral tilt as the higher-level prosodic features. 80-dimensional Mel-spectrograms are computed from pre-emphasized speech using short-time Fourier transform (STFT) with 25~ms frame length and 10~ms shift. The encoder of all the parallel models (4--5) has 4 FFT layers each with a self attention layer having 2 attention heads and 256 hidden units, and two 1-D convolution layers each having a kernel size of 9 and 1024 filters. The decoder has 2 dilated convolution blocks with six 1-D convolution layers with dilation rates of 1, 2, 4, 8, 16 and 32, respectively, kernel size of 3, and 256 filters. The feature predictors have two 1-D convolution layers each having a kernel size of 3 and 256 filters. The rate for the dropout layers and the $\epsilon$ value for the layer normalization were set to 0.2 and $10^{-6}$, respectively. We train all the parallel models (4--5) for 300k steps using 16 GPUs and a batch size of 512. All neural models (3--5) use the same back-end WaveRNN model \cite{appleneuraltts2021asru} to generate speech from the Mel-spectrograms, trained separately for each speaker.

\subsection{Objective measures}
\label{sec:objective}

To measure how well the model can control each prosodic dimension, we synthesized speech at different points in the $[-1,1]$ scale using 300 sentences of general text and responses typical to a voice assistant. By varying each dimension independently, we synthesized 24,900\footnote{24,900 $=$ 5 dimensions $\times$ 9 values $\times$ 300 samples $\times$ 2 speakers $+$ 300 baseline samples $-$ (5$-$1) $\times$ 2 $\times$ 300 samples repeated at 0 bias.} utterances, from which we extracted acoustic features and measured how well they reflect the given target prosodic bias.

\begin{figure}[t]
  \centering
  \includegraphics[width=1.0\linewidth]{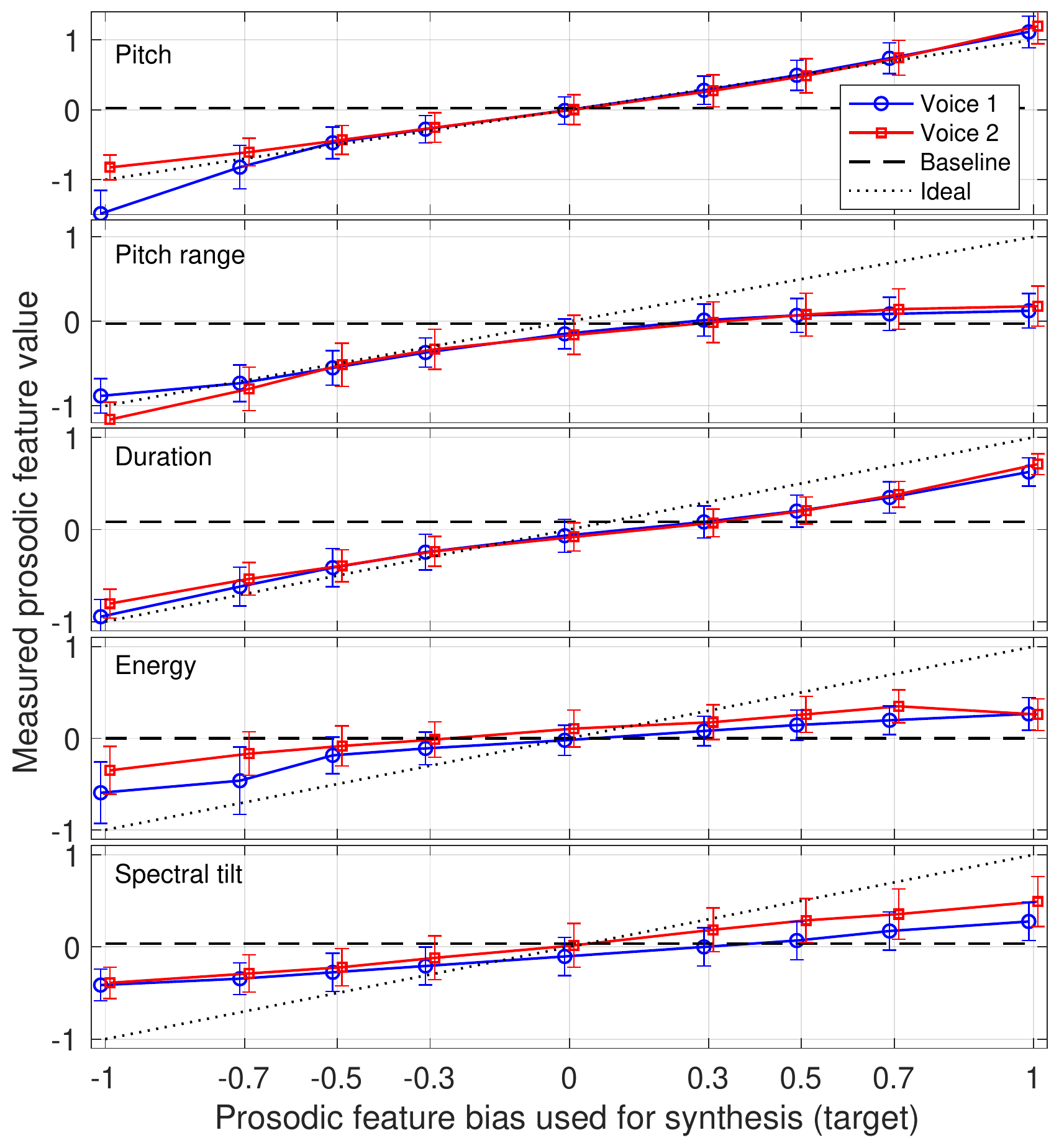}
  \vspace{-8mm}
  \caption{Means and standard deviations of the measured utterance-wise prosodic features with respect to target bias values.}
  \label{fig:measurements}
  \vspace{-4mm}
\end{figure}

\begin{figure}[t]
  \centering
  \includegraphics[width=1.0\linewidth]{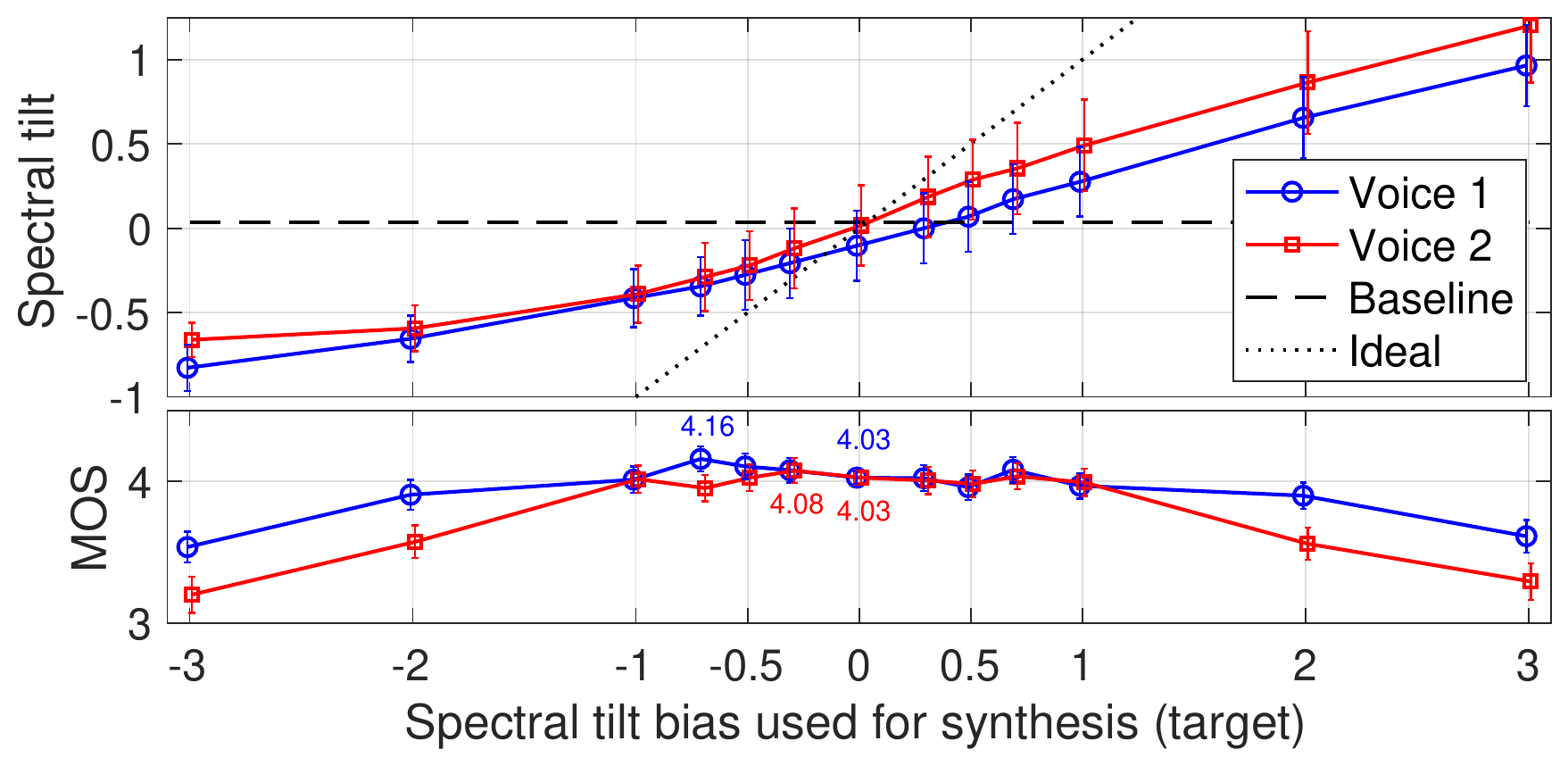}
  \vspace{-7mm}
  \caption{Means and standard deviations of the measured spectral tilt (top) and MOS scores of the synthetic samples (bottom) with respect to target bias values with extrapolation.}
  \label{fig:measurements_tilt}
  \vspace{-4mm}
\end{figure}

Fig.~\ref{fig:measurements} shows the measured acoustic features at each target bias value, normalized by the procedure in Sec.~\ref{sec:prosodic_features} using the original normalization values calculated over the whole database. The original acoustic feature values with respect to the scale $[-1,1]$ are shown in Table~\ref{tab:feature_values}. The results show that the target bias values of the prosodic features are reflected in the output synthetic speech. Although the correspondence between the target bias and the realized prosody does not always match the ideal curve, the modification using the bias enables a significant and easily perceivable change in the corresponding prosody. Moreover, if a stronger change in prosody is required, the model also works when extrapolating the prosodic bias values outside the learned range of $[-1,1]$. This is demonstrated in Fig.~\ref{fig:measurements_tilt} where the spectral tilt bias is varied in the range of $[-3,3]$. Despite not having a linear effect on the output, a stronger prosodic change is achieved with high quality \cite{raitio2021samples}.

\begin{table}[t]
  \caption{Prosodic feature values in normalized scale  $[-1,1]$.}
  \label{tab:feature_values}
  \centering
  \begin{tabular}{|l|l|l|l|l|l|}
    \hline
    Speaker & Feature & Unit & $-$1.0 & 0.0 & 1.0 \\
    \hline
    Voice 1 & Pitch & Hz & 167.0 & 220.2 & 273.4 \\
    Voice 1 & Pitch range & Hz & 95.7 & 294.3 & 492.9 \\
    Voice 1 & Duration & ms & 41.6 & 118.6 & 195.5 \\
    Voice 1 & Energy & dB & $-$24.6 & $-$20.3 & $-$15.9 \\
    Voice 1 & Spectral tilt & - & $-$0.984 & $-$0.955 & $-$0.926\\
    \hline
    Voice 2 & Pitch & Hz & 91.7 & 130.6& 169.5 \\
    Voice 2 & Pitch range & Hz & 77.3 & 230.3 & 383.3 \\
    Voice 2 & Duration & ms & 36.5 & 128.6 & 220.7 \\
    Voice 2 & Energy & dB & $-$23.8 & $-$20.0 & $-$16.1 \\
    Voice 2 & Spectral tilt & - & $-$0.990 & $-$0.961 & $-$0.931\\
    \hline
  \end{tabular}
  \vspace{-4mm}
\end{table}

\subsection{Listening tests}

We carried out a number of listening tests to evaluate the subjective performance of the models. First, we evaluated the overall naturalness of all the models for both speakers. For the proposed model, no prosody bias was applied in this test. We used the 300 synthetic utterances described in Sec.~\ref{sec:objective} for models 2--5 described in Sec.~\ref{sec:models} and 75 natural utterances for both speakers. In addition, we used 150 synthetic utterances for model 1 (with a different speaker). A 5-point mean opinion score (MOS) test was performed by 71 individual American English native speakers using headphones, giving 15 ratings per utterance, resulting in a total of 49,500 responses. The results in Table~\ref{tab:mos_ab} show the following general trend: natural $>$ neural $>$ unit selection $>$ compact. All the neural models yield a high MOS around ~4. The proposed {\it HParallel} model has a significantly higher MOS than the baseline {\it Parallel} model for Voice 1 ($p < 0.003$), while there is no significant difference for Voice 2 ($p = 0.062$).

\begin{table}[t]
  \caption{MOS test results with 95\% confidence intervals computed from the $t$-distribution (upper) and AB test results (lower).}
  \label{tab:mos_ab}
  \centering
  \begin{tabular}{|l|l|l|l|}
    \hline
    \textbf{Test} & \textbf{Model} & \textbf{Voice 1}   & \textbf{Voice 2}  \\
    \hline
    MOS           & Compact         & 1.88 ($\pm$ 0.04) & -                  \\
    MOS           & UnitSelection   & 3.26 ($\pm$ 0.03) & 3.48 ($\pm$ 0.03) \\
    MOS           & Tacotron2       & 4.04 ($\pm$ 0.02) & 3.97 ($\pm$ 0.02) \\
    MOS           & Parallel        & 4.00 ($\pm$ 0.02) & 4.04 ($\pm$ 0.02) \\
    MOS           & HParallel       & {\bf 4.05} ($\pm$ 0.02) & {\bf 4.06} ($\pm$ 0.02) \\
    MOS           & Natural         & 4.24 ($\pm$ 0.04) & 4.32 ($\pm$ 0.04) \\
    \hline
    AB            & Parallel        & 44.8\%             & {\bf 42.5\%}       \\
    AB            & HParallel       & {\bf 46.3\%}       & 42.3\%             \\
    AB            & No preference   & 8.9\%              & 15.2\%             \\
    \hline
  \end{tabular}
  \vspace{-5mm}
\end{table}

%





We also performed AB listening tests where listeners were presented with a pair of speech samples and asked to choose the one that sounded better (or no preference). We evaluated {\it HParallel} and {\it Parallel} models for both speakers. 52 proficient English speakers using headphones participated in the test with 300 sample pairs for each voice with 10 ratings per sample pair, resulting in 6,000 total ratings. The results in Table~\ref{tab:mos_ab} suggest a slight preference for the proposed model for Voice 1 and a slight preference for the baseline for Voice 2, however, the differences are not statistically significant.

\begin{figure}[t]
  \centering
  \vspace{-0.8mm}
  \includegraphics[width=1.0\linewidth]{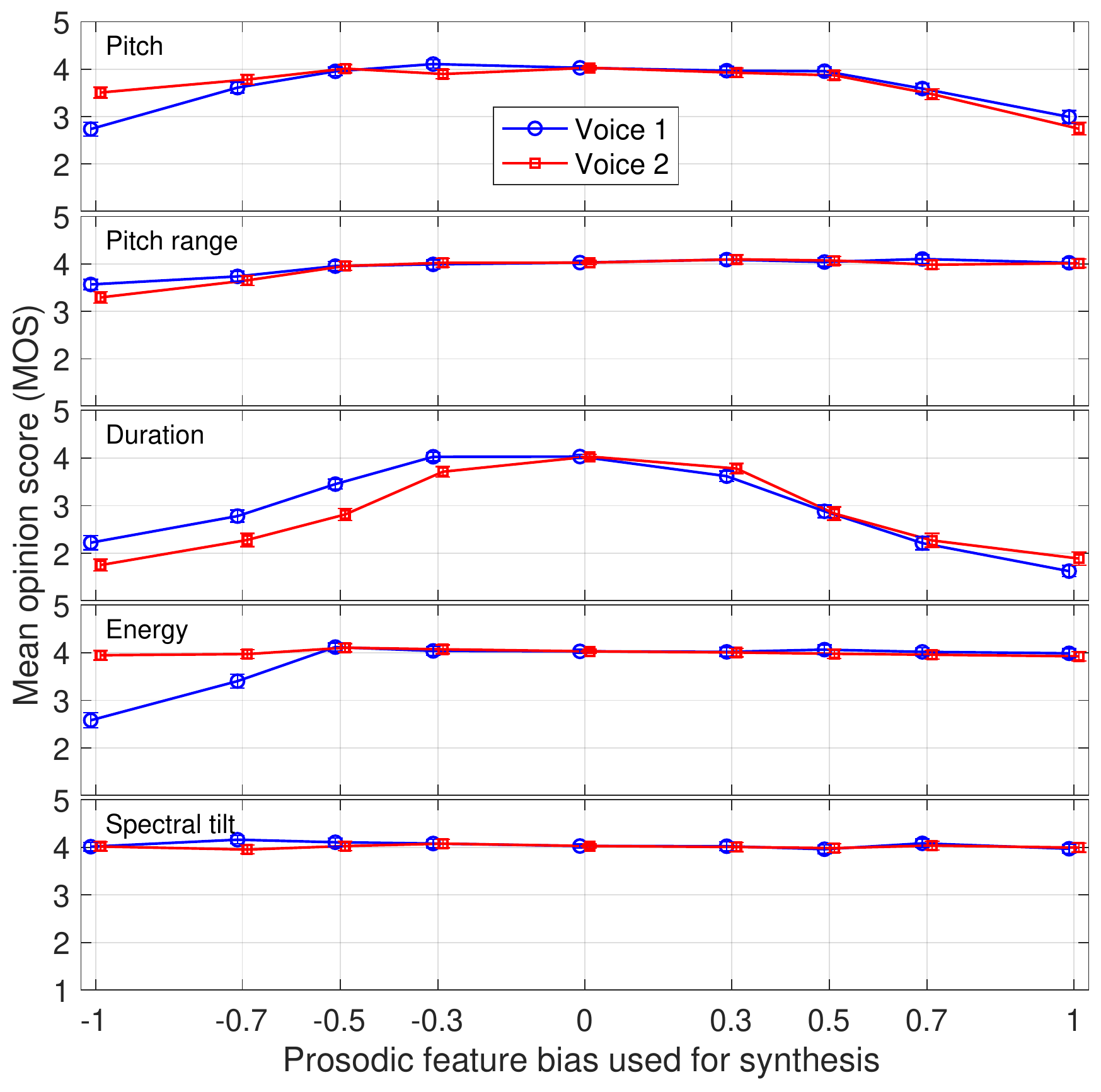}
  \vspace{-7mm}
  \caption{Means and 95\% confidence intervals of MOS measured over all features and target bias values.}
  \label{fig:mos}
  \vspace{-3mm}
\end{figure}

We also evaluated the MOS of the prosody control models at different points in the $[-1,1]$ feature scale. We used 300 sentences described in Sec.~\ref{sec:objective}. A total of 24,900 utterances were synthesized, and each was given 1 rating, totaling 24,900 individual ratings by 34 American English native speakers using headphones. The results in Fig.~\ref{fig:mos} show that the quality remains high with meaningful modification range, however, very low or high phone duration over the whole utterance results in a clearly lower quality.

Since our speech data is recorded in a normal speaking style, it is lacking the full range of spectral tilt variation in human speech. To create more extreme speaking styles, we extrapolated spectral tilt in the range of $[-3,3]$, and then measured the MOS of speech samples. The results in Fig.~\ref{fig:measurements_tilt} show that a much larger change in spectral tilt is observed while the quality remains high even with high degree of extrapolation. Interestingly, the peak MOS values are observed at spectral tilt bias values of $-0.7$ (MOS 4.16) and $-0.3$ (MOS 4.08) for Voices 1 and 2, respectively. This indicates that softer voices may be preferred over the default speaking style. The best way to demonstrate the expressive capability of our proposed model is by listening. We present a set of synthetic speech samples in \cite{raitio2021samples}.

\subsection{Emphasis control}

Another benefit of the proposed approach is that, even if the high-level prosodic features are trained at the utterance level, the model still shows good word-level prosody control ability. Control of emphasis (also lexical focus or prominence) is important in TTS \cite{malisz19_ssw, Yasuda2019, Suni_2020, Shechtman2021} to convey information about which part of the sentence contributes new, non-derivable, or contrastive information. To demonstrate this ability, we synthesized 192 utterances with one word emphasized at a time by adding a bias value of 0.5 to pitch range and duration for the phones in the word. Then we asked 17 listeners to rate the degree of emphasis of the highlighted words on a 4-point scale: 1-neutral, 2-slightly emphasized, 3-emphasized, 4-strongly emphasized. The results with 3,795 ratings show that emphasis is increased on average from 2.44 to 2.66 in comparison to non-biased utterances. We also assessed the AB preference of the utterances, and the results with 1,895 ratings show 51.6\% preference for the emphasized utterances over 30.5\% for the non-emphasized ones (17.9\% no pref). Speech samples on emphasis control are available in \cite{raitio2021samples}.

\section{Conclusions}

We proposed a hierarchical non-autoregressive parallel neural TTS front-end model with prosody modeling and control using intuitive prosodic features. Subjective results show that the proposed model can synthesize speech with equal or better quality to our baseline while being able to control the prosody at utterance level to generate various speaking styles and control emphasis at word level.

\vfill\pagebreak
\bibliographystyle{IEEEbib}
{\ninept\bibliography{mybib}}

\end{document}